\documentclass[conference]{IEEEtran}
\IEEEoverridecommandlockouts
\usepackage{cite}
\usepackage{amsmath,amssymb,amsfonts}
\usepackage{graphicx}
\usepackage{textcomp}
\usepackage{xcolor}
\def\BibTeX{{\rm B\kern-.05em{\sc i\kern-.025em b}\kern-.08em
    T\kern-.1667em\lower.7ex\hbox{E}\kern-.125emX}}

\usepackage{balance}
\usepackage{booktabs} 
\usepackage{subcaption}
\usepackage{cleveref}
\usepackage{multirow}
\usepackage{breqn}
\usepackage{algorithm}
\makeatletter
\def\BState{\State\hskip-\ALG@thistlm}
\makeatother

\usepackage[noend]{algpseudocode}
\newcommand{\cc}[1]{\textcolor{black}{#1}}

\begin{document}

\title{InfDetect: a Large Scale Graph-based Fraud Detection System for E-Commerce Insurance}

\author{
\centering

\IEEEauthorblockN{Cen Chen}
\IEEEauthorblockA{
\textit{Ant Financial Services Group}\\
Hangzhou, China \\
chencen.cc@antfin.com}

\and
\IEEEauthorblockN{Chen Liang\textsuperscript{*}\thanks{* Equal contribution}}
\IEEEauthorblockA{
\textit{Ant Financial Services Group}\\
Hangzhou, China \\
lc155190@antfin.com}

\and
\IEEEauthorblockN{Jianbin Lin}
\IEEEauthorblockA{
\textit{Ant Financial Services Group}\\
Hangzhou, China \\
jianbin.ljb@antfin.com}

\and
\IEEEauthorblockN{Li Wang}
\IEEEauthorblockA{
\textit{Ant Financial Services Group}\\
Hangzhou, China \\
raymond.wangl@antfin.com}

\and
\IEEEauthorblockN{Ziqi Liu}
\IEEEauthorblockA{
\textit{Ant Financial Services Group}\\
Hangzhou, China \\
ziqiliu@antfin.com}

\and
\IEEEauthorblockN{Xinxing Yang}
\IEEEauthorblockA{
\textit{Ant Financial Services Group}\\
Hangzhou, China \\
xinxing.yangxx@antfin.com}

\and
\IEEEauthorblockN{Xiukun Wang}
\IEEEauthorblockA{
	\textit{Ant Financial Services Group}\\
	Hangzhou, China \\
	xiukun.wxk@alibaba-inc.com}

\and
\IEEEauthorblockN{Jun Zhou}
\IEEEauthorblockA{
\textit{Ant Financial Services Group}\\
Hangzhou, China \\
jun.zhoujun@antfin.com}

\and
\IEEEauthorblockN{Yang Shuang}
\IEEEauthorblockA{
\textit{Ant Financial Services Group}\\
San Francisco, USA \\
shuang.yang@antfin.coms}

\and
\IEEEauthorblockN{Yuan Qi}
\IEEEauthorblockA{
	\textit{Ant Financial Services Group}\\
	San Francisco, USA \\
	yuan.qi@antfin.com}

}

\IEEEoverridecommandlockouts
\IEEEpubid{\makebox[\columnwidth]{978-1-7281-0858-2/19/\$31.00~\copyright2019 IEEE \hfill} \hspace{\columnsep}\makebox[\columnwidth]{ }}

\maketitle

\IEEEpubidadjcol

\begin{abstract}
The insurance industry has been creating innovative products around the emerging online shopping activities. Such e-commerce insurance is designed to protect buyers from potential risks such as impulse purchases and counterfeits. Fraudulent claims towards online insurance typically involve multiple parties such as buyers, sellers, and express companies, and they could lead to heavy financial losses. In order to uncover the relations behind organized fraudsters and detect fraudulent claims, we developed a large-scale insurance fraud detection system, i.e., InfDetect, which provides interfaces for commonly used graphs, standard data processing procedures, and a uniform graph learning platform. InfDetect is able to process big graphs containing up to 100 millions of nodes and billions of edges.

In this paper, we investigate different graphs to facilitate fraudster mining, such as a device-sharing graph, a transaction graph, a friendship graph, and a buyer-seller graph. These graphs are fed to a uniform graph learning platform containing supervised and unsupervised graph learning algorithms. Cases on widely applied e-commerce insurance are described to demonstrate the usage and capability of our system. InfDetect has successfully detected thousands of fraudulent claims and saved over tens of thousands of dollars daily.
\end{abstract}


\section{Introduction}

When shopping online, buyers face all kinds of risks. They might receive counterfeits when buying luxury bags; the glass bottle package of spirits might be broken during shipment; the food might be sold after the expiration date. Even when the product is undamaged and genuine, one might still want to return it out of various reasons such as shopping regret or suitability issue after using the product. E-commerce insurance is designed to protect buyers from such risks throughout the complete online shopping process by offering compensations for such unsatisfied experience.

Insurance is a contract used to hedge against future risks and potential financial losses. Any risk that can be quantified can potentially be insured in the form of an insurance policy, which states the conditions and scenarios under which the insurer (i.e., insurance company) will compensate the insured (i.e., policyholder/user). The creation of e-commerce insurance provides a trustworthy environment for both online buyers and sellers and greatly facilitates the active usage of our online shopping website.
The \textit{security deposit insurance} and the \textit{return-freight insurance} are the most popular e-commerce insurance products on Taobao\footnote{One of the biggest e-commerce platforms in the world: https://en. wikipedia.org/wiki/Taobao}. 
The security deposit insurance is purchased by sellers to obtain a `trustworthy seller' badge. If products with quality issues are sold by sellers with this badge, buyers could ask for compensations that are paid by the insurer in advance and is reimbursed by the seller later. Thus sellers are free from the funding pressure for freezing a large amount of security deposit and buyers can still get compensation guarantee when they accidentally purchase products with quality issues. The return-freight insurance is purchased by buyers to protect their right to regret. The insurer pays for the cost of returning unused and undamaged items.

The e-commerce insurance has contributed to over one billion dollars in premiums annually. However, insurance fraud has become a prominent concern. It refers to a range of improper activities that attempt to benefit from a fraudulent outcome from the insurance company~\cite{derrig2002insurance}. 
\cc{According to the estimates of our insurance professionals, millions of potentially fraudulent claims go undiscovered whose costs exceed tens of millions of dollars in each year.} The potential large amount of fraudulent claims could harm both customer satisfaction for the prolonged investigation time and potentially increased premiums, and company's profits, as more human resources and considerable time are required for claim investigations. Thus, it is critical for the insurance company to identify potential fraudulent claims confidently in an efficient manner. The need for a fraud detection system that is able to process very large data arises.

\subsection{Challenges in Insurance Fraud Detection}
Traditional methods on insurance fraud detection primarily focus on extracting handcrafted features (such as past claim history) and subsequently heuristics/rules are distilled based on expert knowledge to decide whether a claim needs further human investigation or not. Witnessing the emergence of big data and distributed computing, insurance companies have started leveraging machine learning techniques to lessen the burden of human investigation/intervention in the claim process~\cite{sithic2013survey}. Statistical models used in insurance fraud detection generally can be categorized into three types: supervised approaches, unsupervised techniques, and a hybrid of both~\cite{joudaki2015using, li2008survey, viaene2002comparison}. Supervised learning approaches, such as logistic regression~\cite{mercer1990fraud, wilson2009analytical}, decision trees~\cite{bonchi1999classification}, support vector machine, Bayesian networks~\cite{ormerod2003using}, and neural networks~\cite{shapiro2002merging, he1997application}, have demonstrated good performances, however, they require data to be labeled by domain experts. On the contrary, unsupervised techniques, such as association rules, cluster analysis, and outlier detection~\cite{brockett2002fraud, yamanishi2004line, viveros1996applying, nian2016auto}, do not have such labeling assumption/limitation and have also attracted much attention over the years. However, there are several aspects that are not well studied in the current literature.

\begin{figure}
\centering
\includegraphics[width=0.34\textwidth]{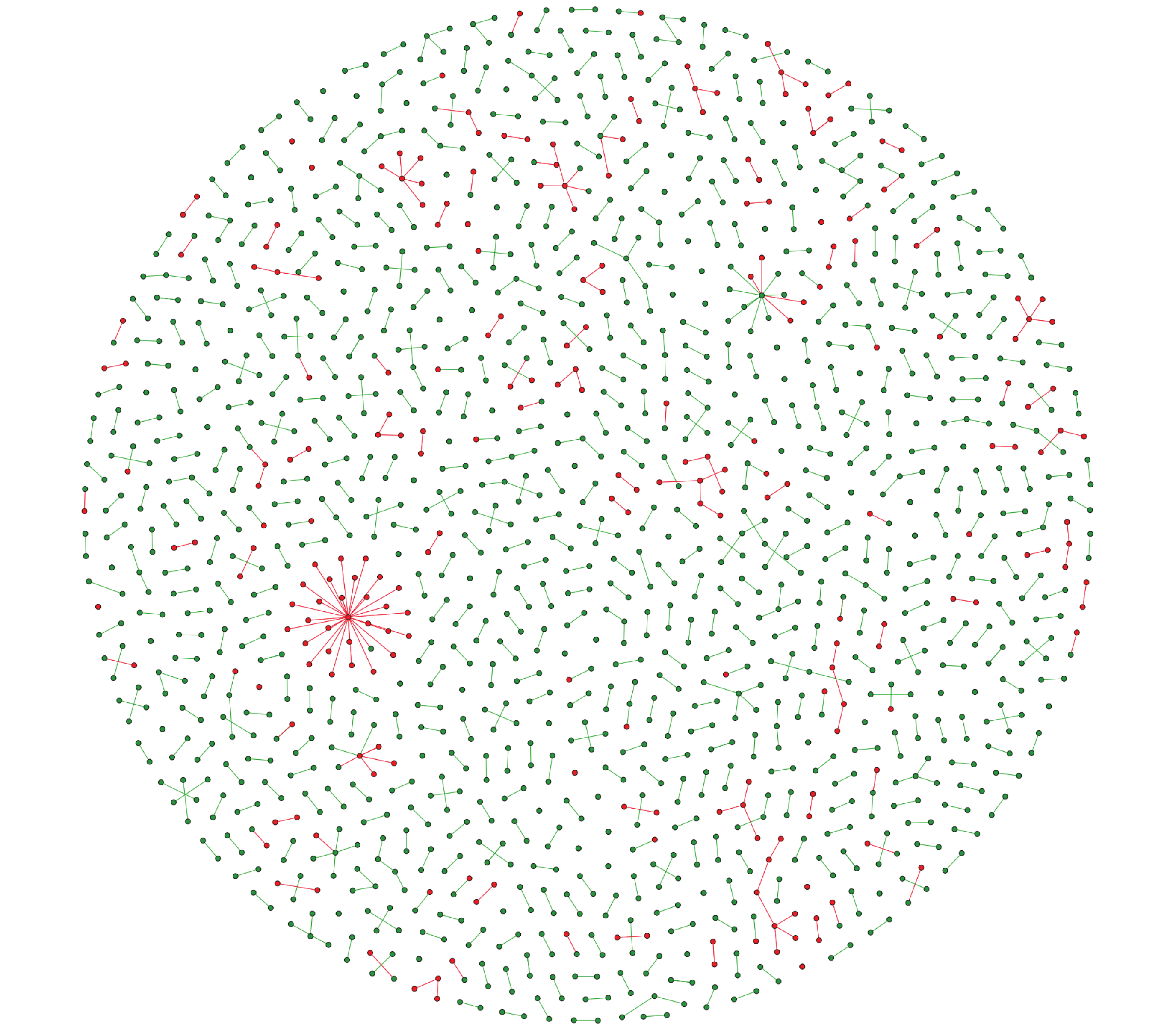}
\caption{Transaction network of a set of sampled claimants
and their neighbors in security deposit insurance, where an edge is formed when there is
a fund exchange between two users. Note that red nodes represent fraudsters while green nodes denote normal users.}
\label{fig:trans}
\end{figure}

\begin{itemize}
   \item \textit{Utilizing both labeled and unlabeled data:} In the insurance domain, it is natural that we have both labeled and unlabeled data. Gathering labels is costly, as long observation period and heavy manual work is often required for labeling. To deal with such problem and boost model performance, one possibility is to combine both supervised and unsupervised learning techniques to better squeeze information from both labeled and unlabeled data for training. We address this problem by introducing unsupervised graph learning algorithms and feature processing techniques in the methodology section.
   \item \textit{Fraud patterns from graphs:} Most deliberate fraudulent behaviors manifest in the form of criminal gangs. Individual behavior can be easier to disguise, but the collective behavior traces can hardly be completely covered up. For example, in Figure~\ref{fig:trans}, we can clearly observe several fraud patterns, where red nodes represent the fraudsters. If we could find a way to utilize additional graph information, e.g., social or transaction networks, it could possibly speed up the claim process and help reduce the fraud rate.
   \item \textit{Uncertain labels:} E-commerce insurance normally issue millions of policies daily and labeling claims requires enormous human effort. A few insurance professionals are not enough for the labeling task, and a common practice here is to ask for a set of rules to separate suspicious and normal. Rules can be applied on the account level, order level, and claim level. A fraudulent score is given and a score higher than the predefined threshold is labeled as `high risk', otherwise `no observable risk'. As we obtain labels for our data, it introduces another problem - label uncertainty. Normally We adjust the threshold so we are confident at `high risk' accounts, but it is unclear whether the `no observable risk' accounts are at risk or not. In other words, the labels we have consist of a small amount of true positive labels and a large amount of unknown labels. To collect labels, we randomly undersample samples from the `no observable risk' samples. This strategy is also explained in the methodology section~\ref{sec:model_uncertainty}.
\end{itemize}

In the rest of the paper, we introduce a large scale fraud detection system for e-commerce insurance that involves all aspects mentioned before. The system is designed to uncover fraudsters in the claim stage by classifying accounts or orders as fraudulent or not. We specifically address the problem of fraudster gang detection with the help of several powerful graph learning algorithms including unsupervised Deepwalk~\cite{perozzi2014deepwalk} and supervised DistRep and 
GeniePath~\cite{liu2018geniepath}. The merits, knowledge, and practices we learn from applying graph data are discussed and we show how we apply them on our most popular real-world large-scale e-commerce insurance products.

\section{Methodology}

Insurance fraud detection can be viewed as a binary classification problem. 
Labels of the claims in the training set are obtained from domain experts and our formerly deployed rule-based system with the confidence of a certain extent. We aim to automatically detect more fraudulent claims while retaining high precision.

Graph, such as social, transaction, and communication networks occur naturally in the insurance fraud settings. They provide straightforward information for describing and modeling complex relations.  Our system involves several types of graphs as data interfaces and provides a variety of machine learning algorithm to mine suspicious fraudsters and orders.

Formally, given a set of a claim $i$'s input feature $\mathbf{x}_i$, and the graphs associated with the claims, our goal is to predict the probability of a claim being fraudulent, i.e., $y_i$.

\subsection{System Overview}
Previous e-commerce insurance fraud detection tasks are conducted by separate insurance data analysts. These professionals come up features through experience and domain knowledge and apply a set of rules on these raw features for fraud detection. Our system is the first graph-based fraud detection system that combines their feature knowledge and various existing graphs. 
\begin{figure}[!htbp]
    \centering
    \includegraphics[width=0.9\columnwidth]{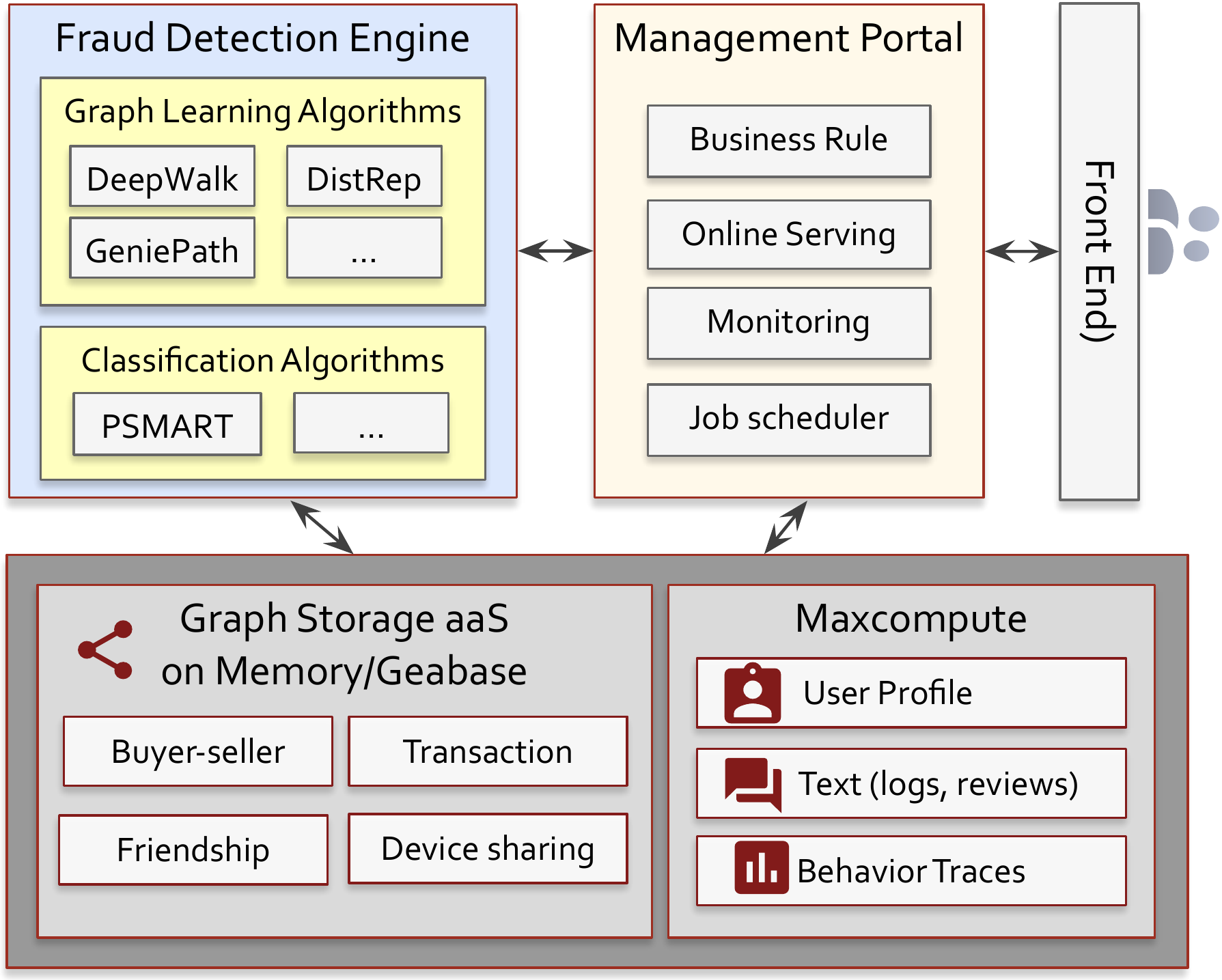}
    \caption{Overview of our insurance fraud detection system.}
    \label{fig:system}
\end{figure}

As shown in Figure~\ref{fig:system}, our system supports two types of algorithms in the fraud detection engine, i.e., graph algorithms to leverage graph information and classification algorithms for general fraud classification widely used in the insurance domain. 
The insurance fraud detection engine is responsible for interacting with the database, model training, and making predictions.
Maxcompute is a general-purpose, fully managed, multi-tenancy data processing platform for large-scale data warehousing~\footnote{Maxcompute: https://www.alibabacloud.com/product/ maxcompute}. It supports SQL and MapReduce for label extraction and feature processing.
At the same time, all the graphs are stored and manipulated in GeaBase~\footnote{GeaBase: https://tech.antfin.com/products/GEABASE}. It is a specially designed graph database used in our company that maintains the n-hop graph neighbor information in a systematic way. It is able to store large graphs with low lookup latency.
Meanwhile, the management portal supports a variety of management tasks across the whole pipeline, such as business rule intervention, online serving, monitoring, and job scheduling.

\subsection{Data Processing}

Features are collected and processed to be fed into downstream machine learning algorithms in more suitable representations. The data processing modules provide several common utility functions such as data scaling, categorical feature encoding, discretization, and missing values filling. 

Aside from basic features processed from the raw inputs, we can further enrich the representation of fraud patterns by incorporating denoised latent feature embeddings, which leverage the \textit{Denoising Autoencoder} (DAE)\footnote{The details of DAE is omitted, as it is not the focus for the paper.} to transform basic features from a corrupted version for robustness and better generalization.
Such unsupervised feature transformation techniques help to better distill additional information from unlabeled data. 

Besides, population stability index (PSI)~\cite{yurdakul2018statistical} is measured to find out whether a feature is significant enough for classification and stable enough along time. It is used to measure how much a variable has shifted in distribution between two samples. Commonly it is used to monitor the distribution changes of a feature between out-of-time validation samples and modeling samples. If the change is significant, this feature is not valid for online production because of stability issues. PSI is also used to decide whether a feature is important in the modeling stage. If the distribution difference is large between positive samples and negative samples, the feature is retained for modeling.

In addition, graph-based features are extensively used as an essential part in our system. From the graph theory perspective, features such as the degree of a node, the index of the subgraph a node resides in, and the length of the longest path containing a node are precomputed. Because our graphs are stored as assets, computing such features in advance could save a great amount of time when shared in every downstream fraud detection tasks. From the representation learning perspective, graph embedding learned by supervised and unsupervised graph learning algorithms can also be incorporated to uncover potential conspiracy patterns. 

Finally, all these features will be concatenated and fed into the classification algorithms.

\subsection{Classification Algorithms}
Different from fraud detection systems in other domains, insurance claimants are rather sensitive and alert to the results.  For models used in the insurance industry, \textit{interpretability} is sometimes one of the most important concerns. For example, for some insurance, when the company rejects a claim, the verifier may have to explain the possible reasons/fraud indicators associated with the claim. 
As a result, classification algorithms with good explainability, such as logistic regression~\cite{wilson2009analytical,mercer1990fraud},
decision trees~\cite{bonchi1999classification}, are often utilized. 
In our system, we have implemented a series of general classification algorithms.
Parameter server based gradient boosted decision trees, also known as PSMART~\cite{zhou2017psmart}, is mostly adopted for its \textit{good expressive power, scalability, and explainability}. 
More specifically, PSMART is distributed implemented over parameter server~\cite{li2014communication} on top of the tree boosting technique LambdaMART~\cite{burges2010ranknet}.
It is deeply optimized for the communication efficiency over the sparse data that can reliably scale to hundreds of billions of samples and thousands of features over the clusters.

\subsection{Graph Learning Algorithms}

To help uncover the collective fraudster traces, we leverage the graph representation learning models to bring additional graph-based latent information into the picture. 
In the following subsections, we will dive into the details of three representative graph learning algorithms, i.e., Deepwalk (unsupervised), Graph Neural Networks (supervised node classification), and DistRep (supervised edge classification). All the algorithms are developed in a distributed fashion over parameter server to handle large scale graphs of up to billions of nodes.

\subsubsection{Deepwalk}
Deepwalk (DW) belongs to the family of unsupervised graph learning models. Such models are able to leverage the unlabeled graph data, capture neighborhood similarity and encode the topological relationships into a latent vector space in the form of \textit{embedding}~\cite{goyal2017graph}. 
DW uses local topological information obtained from truncated random walks sampled from the graph to learn latent representations by treating walks as the equivalent of sentences. 
Following \cite{perozzi2014deepwalk}, the learning procedure in Deepwalk is formulated as a maximum likelihood optimization problem:
\begin{equation}
\max_f \sum_{u\in V} \log Pr\left(N \left(u)\right|f\left(u\right)\right),
\end{equation}
where $f$ is a matrix of size $\left | V \right | \times d$ parameters. For each vertex $u\in{V}$, it defines $N\left ( u \right )\in{V}$ as a network neighborhood of source vertex $u$ generated through the random walk.

\begin{figure}[!htbp]
    \centering
    \includegraphics[width=0.9\columnwidth]{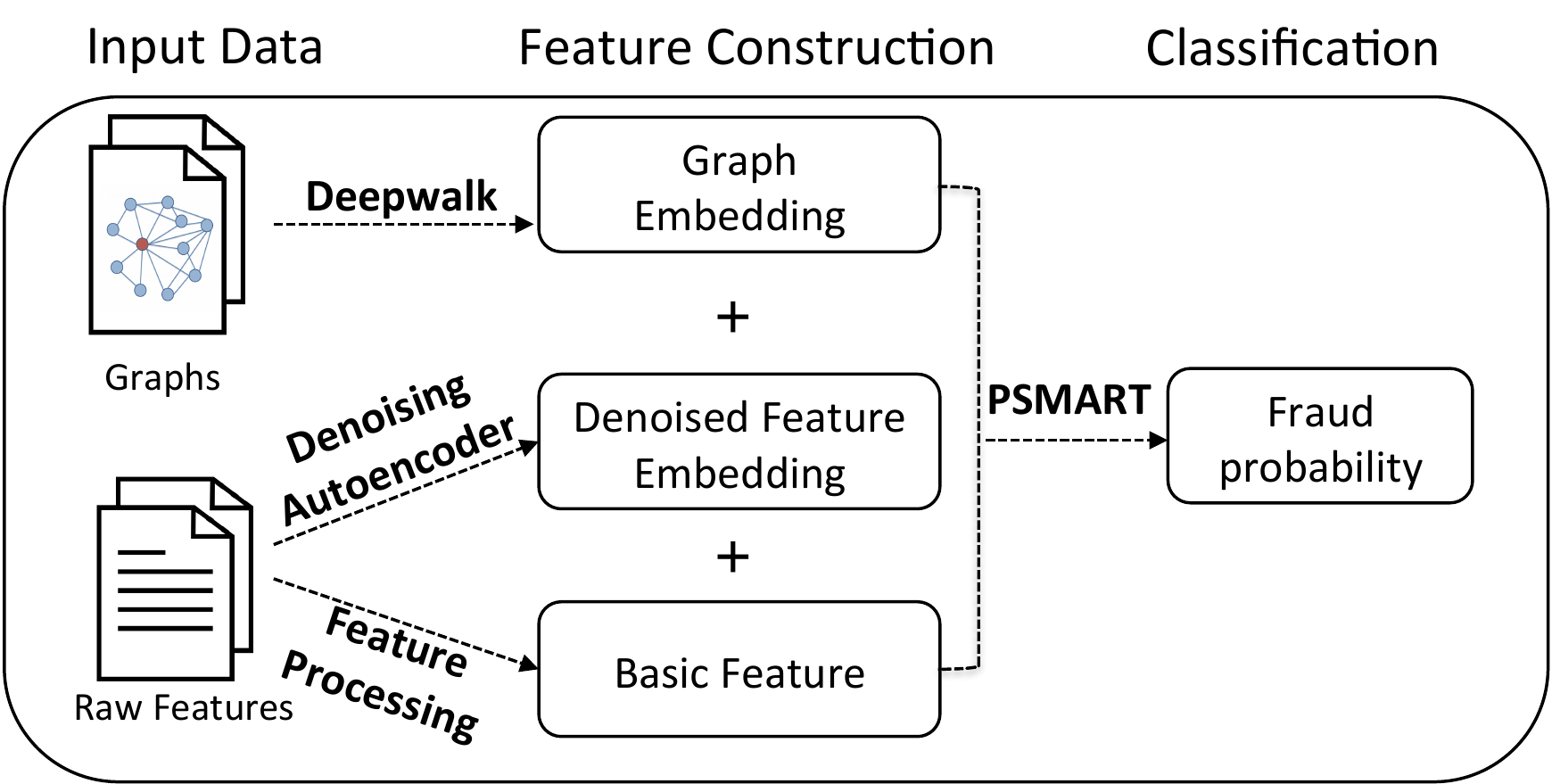}
    \caption{Fraud detection pipeline with graph embedding.}
    \label{fig:pipline}
\end{figure}

For such unsupervised graph learning technique, the learned embeddings usually serve as the input features for the downstream tasks. A common practice for fraud classification with graph embeddings is outlined in Figure~\ref{fig:pipline}.

\subsubsection{Graph Neural Networks (GNNs)}
GNNs are a set of deep learning algorithms following the same architecture that aggregates information
from nodes' neighbors. A deeper layer reaches out more distant neighbors, and the $k$th layer embedding of node $v$ is 
\begin{displaymath}
  \mathbf{h}_v^k = \sigma (\mathbf{W}_k \cdot  \mathrm{AGG}(\mathbf{h}_u^{k-1},\forall u \in \mathcal{N}{(v)} \cup \{v\})) 
\end{displaymath} where the initial embedding $\mathbf{h}_v^0=\mathbf{x}_v$ is the account feature, $\sigma$ is a non-linear function, and $\mathrm{AGG}$ is an aggregation function across layers and neighbors that differs in GNN algorithms.

Common GNN approaches we use for the fraud detection problem are struct2vec~\cite{dai2016discriminative} and GeniePath~\cite{liu2018geniepath}. Struct2vec aggregates neighbors by simply summing them up while GeniePath stacks adaptive path layers for breadth and depth exploration in the graph. For breadth exploration, it aggregates neighbors as
\begin{eqnarray}
  \mathbf{A} = \mathrm{tanh}(\mathbf{W}_s \mathbf{h}_v^{k} + \mathbf{W}_d \mathbf{h}_u^{k}) \notag  \\
  \mathrm{AGG}(\mathbf{h}_u^{k}) =\sum_{u \in \mathbf{N}(v) \cup \{v\}}{\mathrm{softmax}(\mathbf{w^T}\mathbf{A})\cdot \mathbf{h}_u^k} \notag 
\end{eqnarray} 
This breadth-search function emphasizes the importance of neighbors with similar account features.

The resulting embeddings are fed to the final softmax or sigmoid layers for downstream fraud account classification tasks. It's an end-to-end classification method compared to Deepwalk whose embeddings are treated as features to downstream classification algorithms.

\subsubsection{DistRep}
DistRep is a novel algorithm we designed for edge classification. It combines node embeddings and node attributes. The embeddings of a edges' both vertices $u$ and $v$ are aggregated as \begin{displaymath}
  \mathbf{h}_{\mathrm{emb}}^{\{u, v\}} = \mathrm{dropout}(\mathbf{h}_u) + \mathrm{dropout}(\mathbf{h}_v)
\end{displaymath} while the attributes of both vertices are concatenated as  \begin{displaymath}
  \mathbf{h}_{\mathrm{att}}^{\{u, v\}} = \sigma (\mathbf{W}_{\mathrm{att}} \cdot \mathrm{concat}(\mathbf{h}_u^0, \mathbf{h}_v^0))
\end{displaymath} where $\mathbf{h}_v^0$ and $\mathbf{h}_u^0$ are the node features. $\mathbf{h}_{\mathrm{emb}}^{\{u, v\}}$ and $\mathbf{h}_{\mathrm{att}}^{\{u, v\}}$ are concatenated and fed into a $k$-layer neural network. The final sigmoid layer output the edge classification result.

\subsection{Modelling Label Uncertainty}
\label{sec:model_uncertainty}
Most e-commerce datasets suffer from label uncertainty - the rule-based risk indicator is much more confident about `high risk' accounts being fraudulent than about `no observable risk' accounts being regular. To address this problem, the `regular' class in the training dataset is sampled randomly. Downsampling helps to reduce the effect of classifying a `no observable risk' account as fraudulent. The objective function is modified as follows

\begin{displaymath}
\begin{split}
  \mathcal{L}(w) & =  \min_w( \sum_{v \in \mathcal{V}_{\mathrm{fraudulent}}}\ell(f(\mathbf{x}_v;w), \mathrm{fraudulent}) \\
  & + \sum_{v \in \mathrm{sample}(\mathcal{V}_{\mathrm{regular}})}\ell(f(\mathbf{x}_v;w), \mathrm{regular}))
\end{split}
\end{displaymath}
$f$ represents the classification algorithm of our choice. The goal is to minimize the losses caused by wrong classifications. Note the sampling process only exists when selecting samples to be trained. Once the training samples are selected, their neighborhoods (containing 1-hop to 3-hop neighbors in most applications) are not sampled.

\section{Discussion}

\begin{figure}[!htbp]
\centering
\includegraphics[width=0.7\columnwidth]{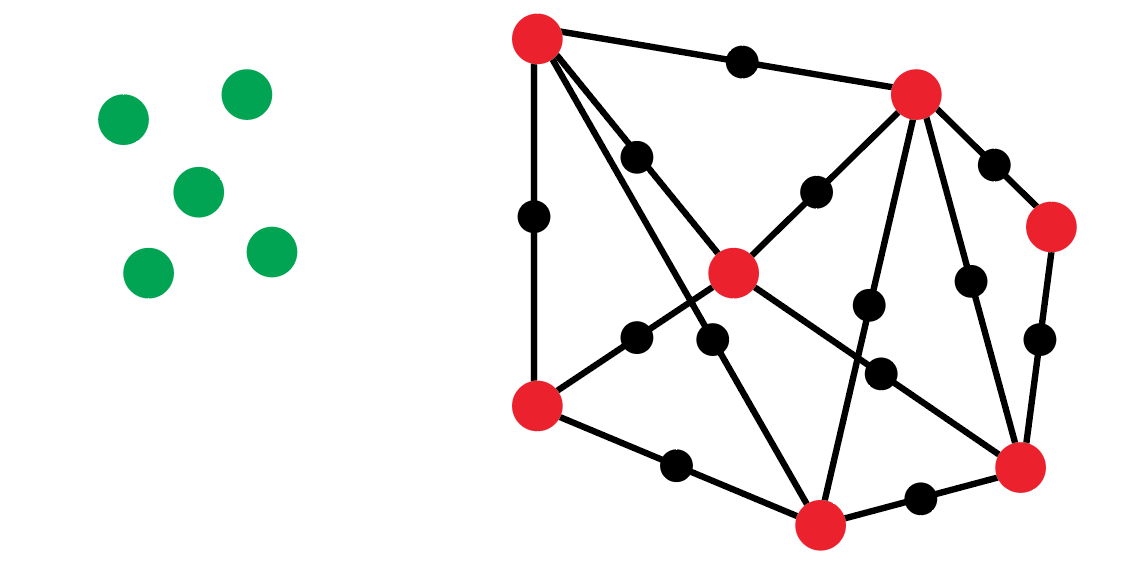}
\caption{Buyer-seller graph of fraudulent users in the order insurance. The red nodes represent fraudsters in sellers and the green nodes denote normal sellers. The larger nodes are sellers and smaller black nodes are buyers. Only essential buyers that connecting sellers are visualized for simplicity.}
\label{fig:orderpattern}
\end{figure}

The key component in InfDetect that differs from other machine learning-based fraud detection systems in the insurance domain is the usage of graph information. Graph is helpful in the following perspectives:

\begin{itemize}
   \item \textit{Fraud Organization Discovery:} As we mentioned in Figure~\ref{fig:trans}, fraudulent accounts are visualized as connective red nodes. In other cases, similar patterns are also discovered (See Figure~\ref{fig:orderpattern}). 
   \item \textit{Fraud Detection with Consistency:} Fraud detection suffers from the phenomenon that new types of fraud evolve over time and get more and more unpredictable. The use of non-stationary features, such as the number of claims made in the past month, can be easily affected when fraudsters change their tactics. Meanwhile, graph data provides more stationary information as the relations between collaborating fraudsters could not be easily modified, e.g., in device-sharing graphs. Thus the use of graphs helps to establish model consistency.
\end{itemize}

\subsection{Graph Construction}

\begin{figure} [!htbp]
  \centering
  \subcaptionbox{Device-sharing: colluders\label{first-subfig}}{%
    \includegraphics[width=0.2\textwidth]{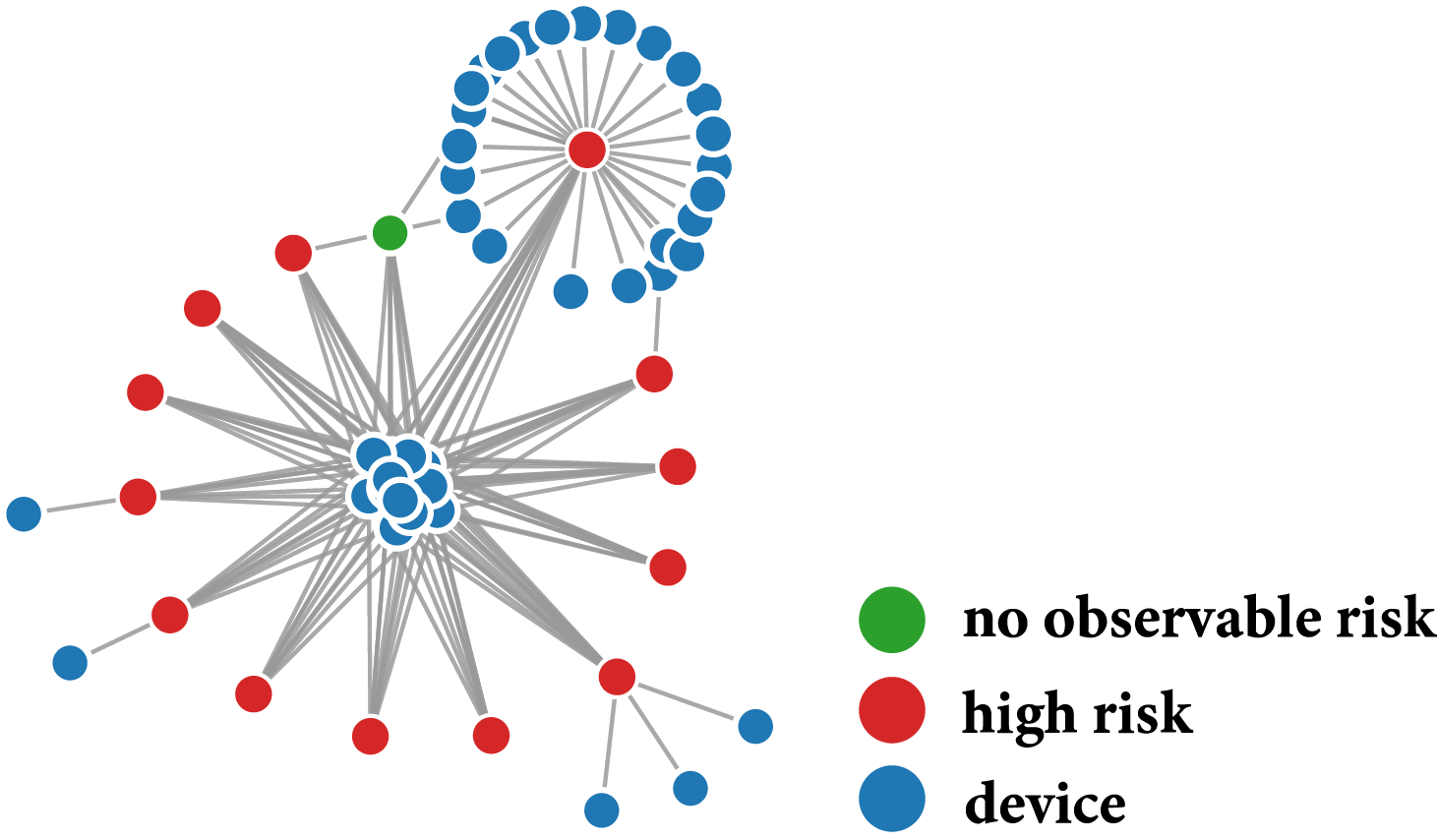}
  }
  \subcaptionbox{Device-sharing: regular\label{second-subfig}}{%
      \makebox[0.2\textwidth][c]{\includegraphics[width=.18\textwidth]
    {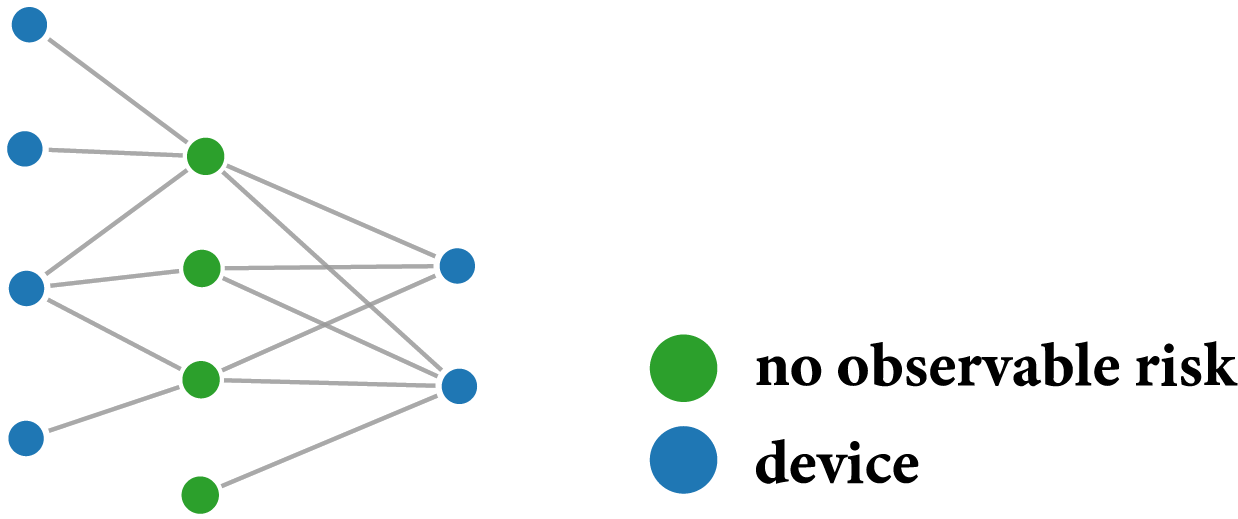}}
  }
  \subcaptionbox{Transaction: colluders\label{first-subfig}}{%
      \makebox[0.2\textwidth][c]{\includegraphics[width=.16\textwidth]
    {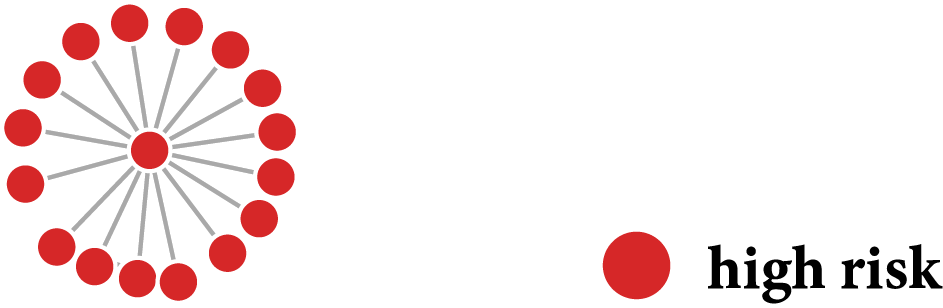}}
  }
  \subcaptionbox{Transaction: regular\label{second-subfig}}{%
    \includegraphics[width=0.2\textwidth,scale=0.4]{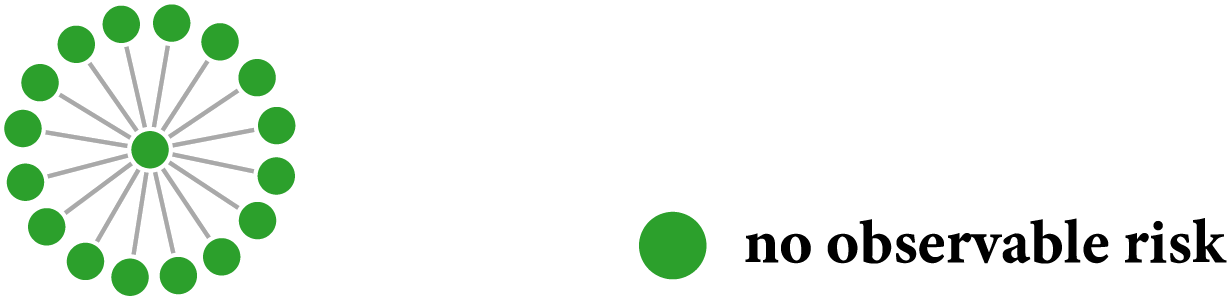}
  }
  \subcaptionbox{Friendship: colluders\label{first-subfig}}{%
    \includegraphics[width=0.2\textwidth]{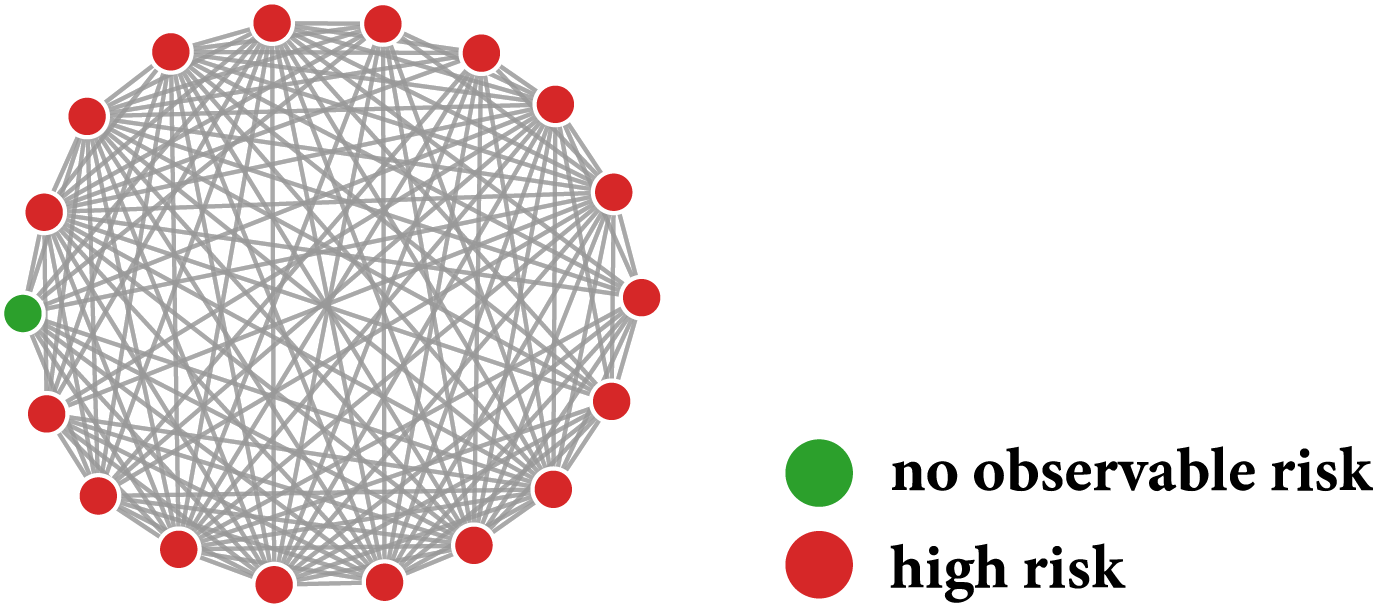}
  }
  \subcaptionbox{Friendship: regular\label{second-subfig}}{%
    \includegraphics[width=0.2\textwidth]{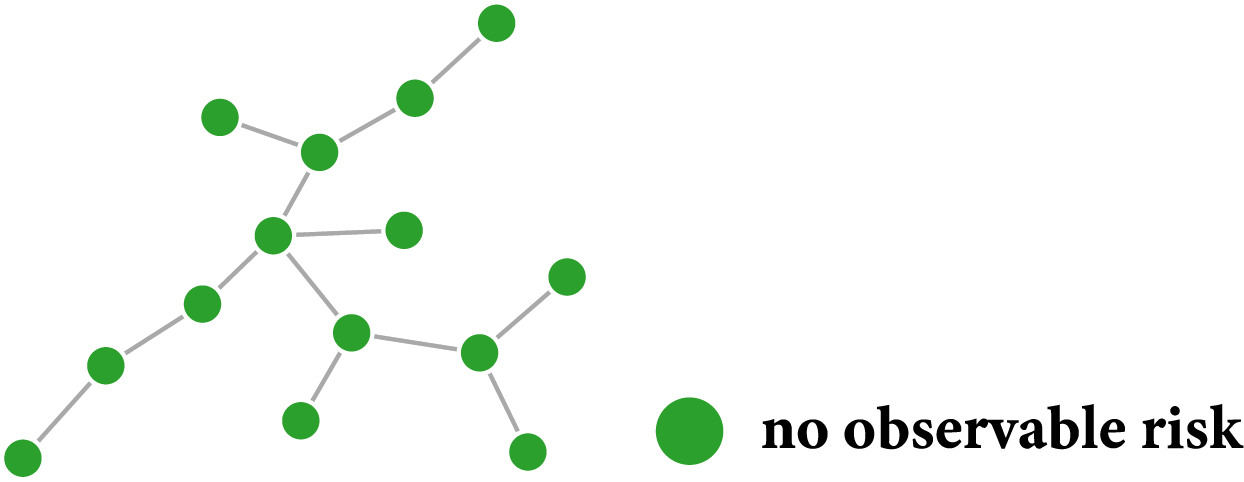}
  }
  \caption{Visualization for typical colluders and regular users in device-sharing graph, transaction graph, and friendship graph.}
  \label{fig:vis}
\end{figure}

In this study, we form the transaction graph, device-sharing graph, and friendship graph to reveal patterns for fraud classification (see Figure~\ref{fig:vis}), and build a buyer-seller graph to identify fraudulent orders. The following properties of graphs can help separate fraudulent from regular:

\begin{itemize}
\item \textit{distance aggregation:} closer nodes share similar labels;
\item \textit{structural differentiation:} structures of organized fraudsters are different from structures of regular accounts.
\end{itemize}

\subsubsection{Buyer-Seller Graph}
The buyer-seller graph is built based on Taobao's order history. Orders from the past week as collected and each edge corresponds to one order while its two vertices corresponds to a seller account and a buyer account, respectively.

\subsubsection{Transaction Graph}

The transaction graph shows fund exchange relations between accounts. A vertex is an account, and an edge indicates transactions between accounts.

\subsubsection{Device-Sharing Graph}

The device-sharing graph reveals the relation of accounts sharing a device. A vertex is either a device (User Machine ID, UMID\footnote{The fingerprint built by Alibaba to uniquely identify devices.}) or an account. Edges exist between a device vertex and a UMID vertex, which are extracted from the log-in history.

\subsubsection{Friendship Graph}

The friendship graph is built upon friend books at Alipay, a product of Ant Financial with social networking features.

\subsection{Graph Processing}

We preprocessed these graphs to remove isolated accounts. In the transaction graph and friendship graph, nodes with zero degree (the number of edges incident to the node) are removed. In the device-sharing graph, account nodes who share no common UMIDs with other accounts and their neighboring UMID nodes are removed.

With the graph processing step, the classification performance is slightly degraded by less than 0.1\%, whereas a great amount of computation is saved - the computation time for DeepWalk is shortened from 45 hours to 8 hours after processing the device-sharing graph.

\subsection{How to Choose Graphs}

The graphs are of great size (see Table~\ref{tag:stat}), and we evaluate the graphs in advance to avoid implementing all graphs at hand for efficiency. The evaluation metrics are designed in regards to the distance aggregation policy which states if closer nodes in a graph have similar labels, this graph is more helpful for this classification task. We measure it by:
\begin{displaymath}
  \eta = \max_{hop\in \{\mathbf{1, 2, ...,H}\}}\frac{\sum_{i \in B}\left | N_B^{hop}(i) \right |}{\sum_{i \in \{B,W\}}\left | N^{hop}(i) \right |},
\end{displaymath}
where $B$ is the set of fraudulent nodes and $W$ is the set of normal nodes.

\begin{table*}[!htbp]
\centering
  \label{tab:graph-info}
  \begin{tabular}{ccccc}
  \toprule
    \textbf{Graph} & |V| & |E| & nodes & edges \\
    \midrule
    device-sharing & 3 M & 6 M & account / UMID & device usage \\ 
    transaction & 2 M & 2 M & account & fund exchange \\
    friendship & 8 M & 11 M & account & friendship \\
    buyer-seller & 100 M & 1 B & account & product purchase \\
  \bottomrule
\end{tabular}
\caption{Examples of the Graphs provided in InfDetect. $V$ and $E$ denote the vertices and edges, respectively.}
\label{tag:stat}
\end{table*}

\subsection{How to Use and Choose Graph Learning Algorithms}

Graph information can be used as features in traditional machine learning algorithms. One example is to compute the in-degree and out-degree of a node. Graph knowledge is partially considered in a simple but powerful way, and in some cases, it can lead to a slight performance improvement. For example, when the fraudsters are working with a so-called `mobile phone factory'\footnote{A large amount of inexpensive mobile phones are purchased by fraudsters to register fake accounts and conduct fraud.}, degree of fraudster account nodes in the device-sharing graph is significantly higher than others. 

The usage of graph information as features is not as powerful when attempting to discover relations between certain fraudsters where graph learning algorithms are preferred. In the case of order-wise fraud detection, DistRep is more appropriate as it considers an order as an edge between a seller and a buyer. As for account-level fraud detection, graph neural networks work end-to-end and the embeddings extracted from its hidden layers are task-specific and contain label information. Meanwhile, DeepWalk distills graph structural information and gives a set of uniform embeddings of nodes regardless of downstream tasks.

\section{Cases study on E-Commerce Insurance}
In this section, we quantitatively and qualitatively evaluate the effectiveness of our graph-based fraud detection system over our mainstream products of e-commerce insurance.

\subsection{Security Deposit Insurance}
Security Deposit Insurance is one of the most popular insurance for sellers on Taobao.
To obtain a `trustworthy seller' badge, a seller can choose to freeze a security deposit fund or to buy the security deposit insurance with a yearly premium of a small amount.  Such insurance helps the insurer to pay for the emergency compensation in advance.

\subsubsection{Data Preparation}
Our security deposit insurance dataset is sampled from its claim history 
One transaction graph is generated for each day. More specifically, for users (sellers/ buyers) involved in the claims on a day, we retrieve their corresponding transaction records from our platform to build a transaction graph. 
On average, each transaction graph contains $500k$ nodes and  $800k$ edges.

\subsubsection{Quantitative Evaluation}

We conduct ablation experiments to examine the effectiveness of incorporating the graph information, i.e., embedding learned by DeepWalk (DW). Our parameter server based GBDT method--PSMART~\cite{zhou2017psmart} is used as the base classification model. 
Grid search is performed to find the best parameter settings. 
Both graph embedding size for DW and denoised feature embedding size for DAE are set as 32.
As shown in Table~\ref{tab:features}, incorporating DW significantly boost the model performance. Both DAE and DW are helpful for the task.
\begin{table}[!h]
    \centering
    \begin{tabular}{l|c |c |c }
        \toprule
         & AUC &Rec.@90\%Pre.& Rec.@95\%Pre. \\
        \midrule
        PSMART   &0.9650 &44.71\% &69.30\%   \\
        PSMART+DAE   &0.9655 &46.48\% &71.04\%   \\
        PSMART+DW   &0.9661 &46.75\% &74.49\%   \\
        PSMART+DAE+DW  &\textbf{0.9667} &\textbf{47.12\%} &\textbf{77.89\%} \\
        \bottomrule
    \end{tabular}
    \caption{
    Performance comparison in terms of AUC and Recall (Rec.) at different Precision (Pre.) thesholds.
    }
    \label{tab:features}
\end{table} 

\subsubsection{Online Performance}
After an A/B test for 1 month on our platform, we find that our proposed method is able to reduce fraud rate by 76\% compared to the previous rule-based method
\footnote{We cannot report the accurate insurance claim amount due to the privacy issue.}. 

\subsubsection{Qualitative Evaluation}
To understand why our model has better performance on insurance fraud detection task, we qualitatively evaluate our method from two perspectives: one is at claim-level and the other is at user-level. More specifically, we visualize the learned graph embeddings of DW using the t-SNE tool\footnote{T-SNE is a commonly used tool for the visualization of high dimensional data.}.

\begin{figure}[!htbp]
    \centering
    \includegraphics[width=0.9\columnwidth]{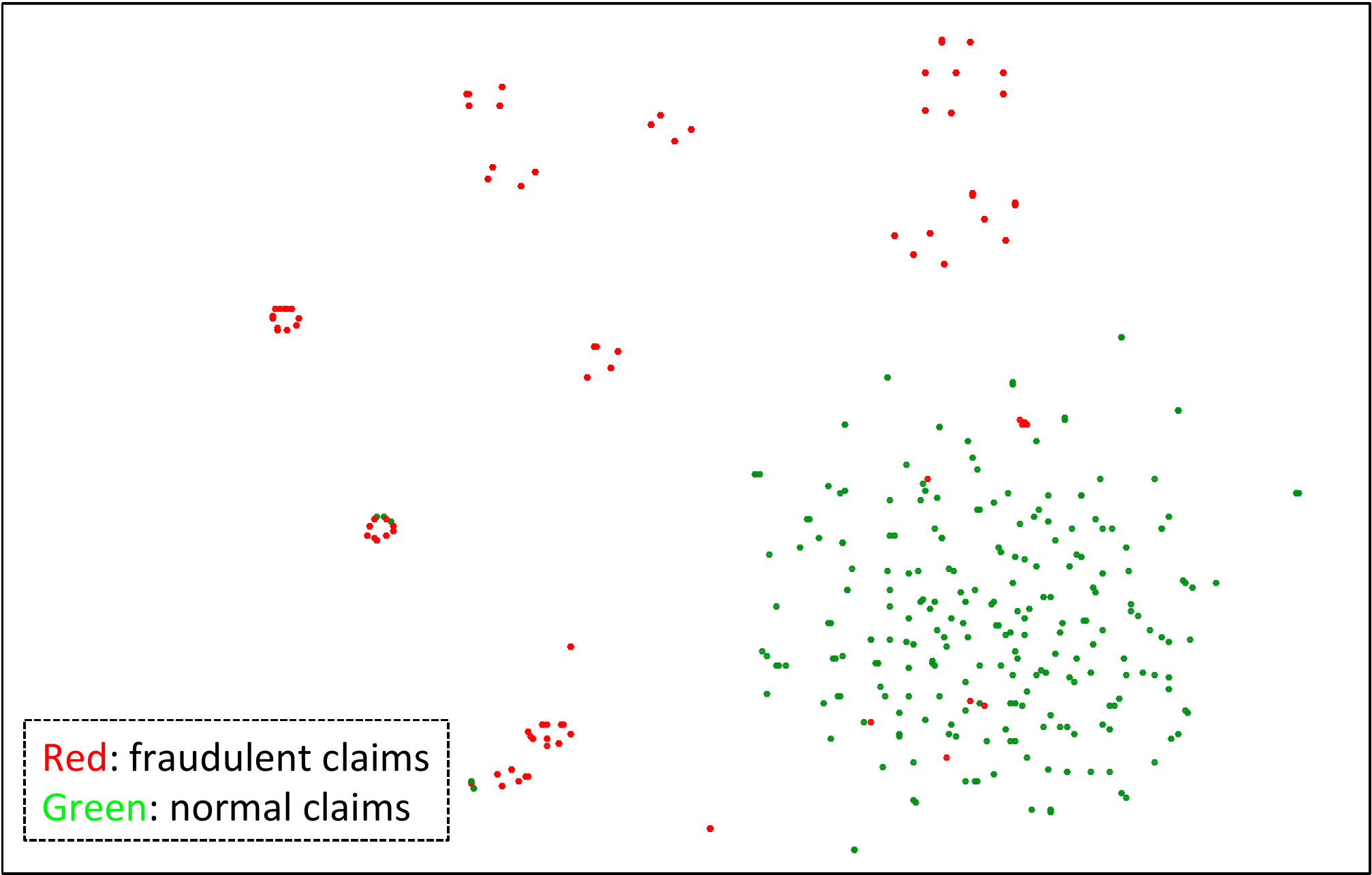}
    \caption{Claim level embedding. Red dots represent the fraudulent claims, while  while greed dots refer to the normal claims.}
    \label{fig:claim_level}
\end{figure}

\textbf{Claim level embeddings}: 
For this particular insurance product, each claim involves two parties, we obtain the claim representations by concatenating the involved user embeddings. We then visualize the sampled claims on a day by their representations in Figure~\ref{fig:claim_level}. Clearly, we find fraudulent claims (in red) are not close to the normal claims (in green). This shows the graph representations are useful for identifying fraudulent claims. Furthermore, we observe that the fraudulent claims form different small clusters. This demonstrates that there is a gang behavior in the fraudulent claims, i.e. there are small groups of users colluding on insurance claim fraud together. This further shows the graph representations are meaningful.

\begin{figure}[!htbp]
    \centering
    \includegraphics[width=0.9\columnwidth]{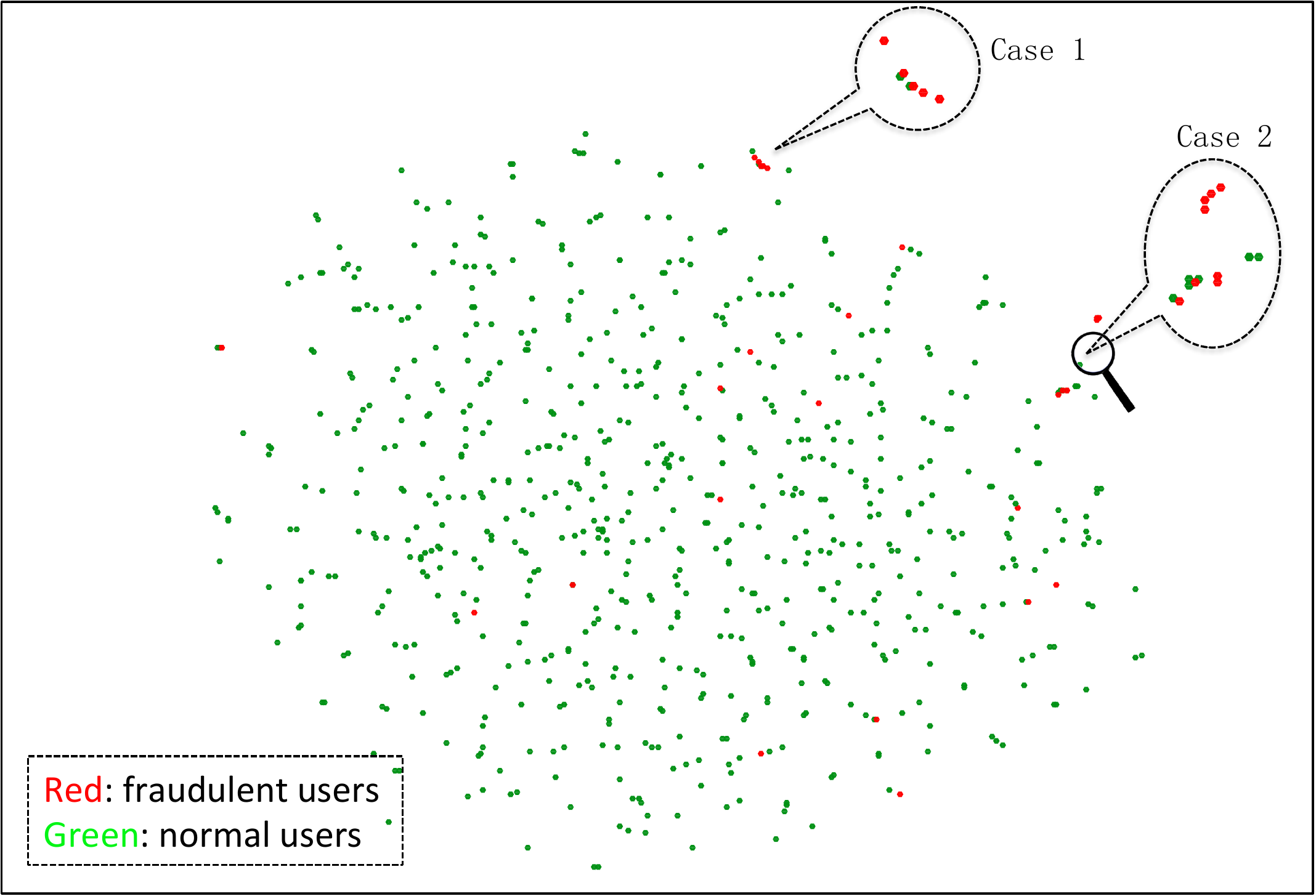}
    \caption{User level embedding. A fraudulent user is a seller or buyer who is involved in a fraudulent insurance claim.}
    \label{fig:user_level}
\end{figure}

\textbf{User embeddings}: Moreover, we visualize the user embeddings learned by our method in Figure~\ref{fig:user_level}. We use red color to mark a fraudulent user who initiated a fraud claim, and green color to mark normal users. Close examination shows that there are small clusters of fraudulent users and our method is able to project the fraudulent users into similar places in the embedding space.

\begin{figure}[!htbp]
    \centering
  \subcaptionbox{Case 1 \label{first-subfig}}{%
    \includegraphics[width=0.24\textwidth]{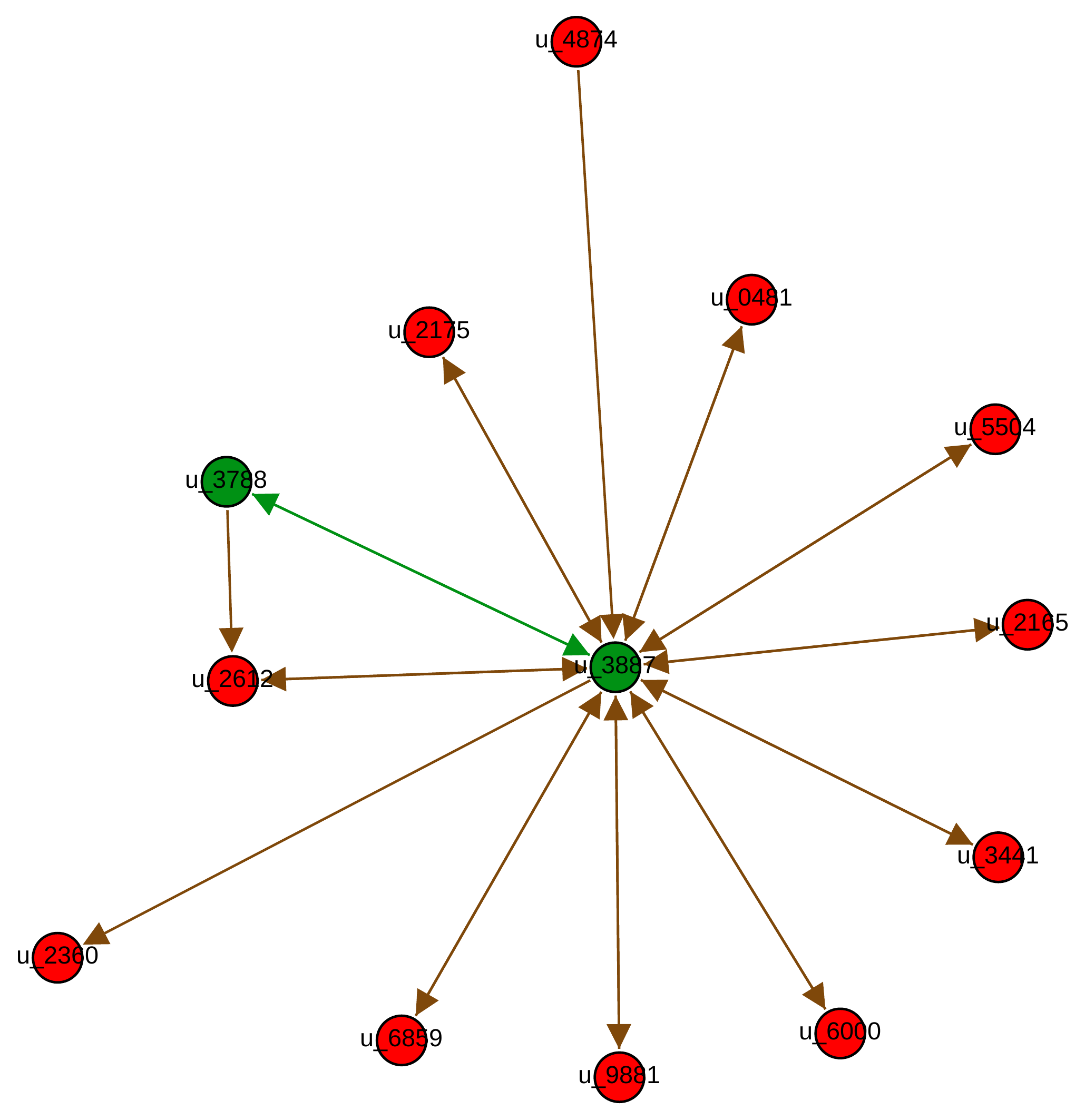}
  }
  \subcaptionbox{Case 2 \label{second-subfig}}{%
    \includegraphics[width=0.18\textwidth]{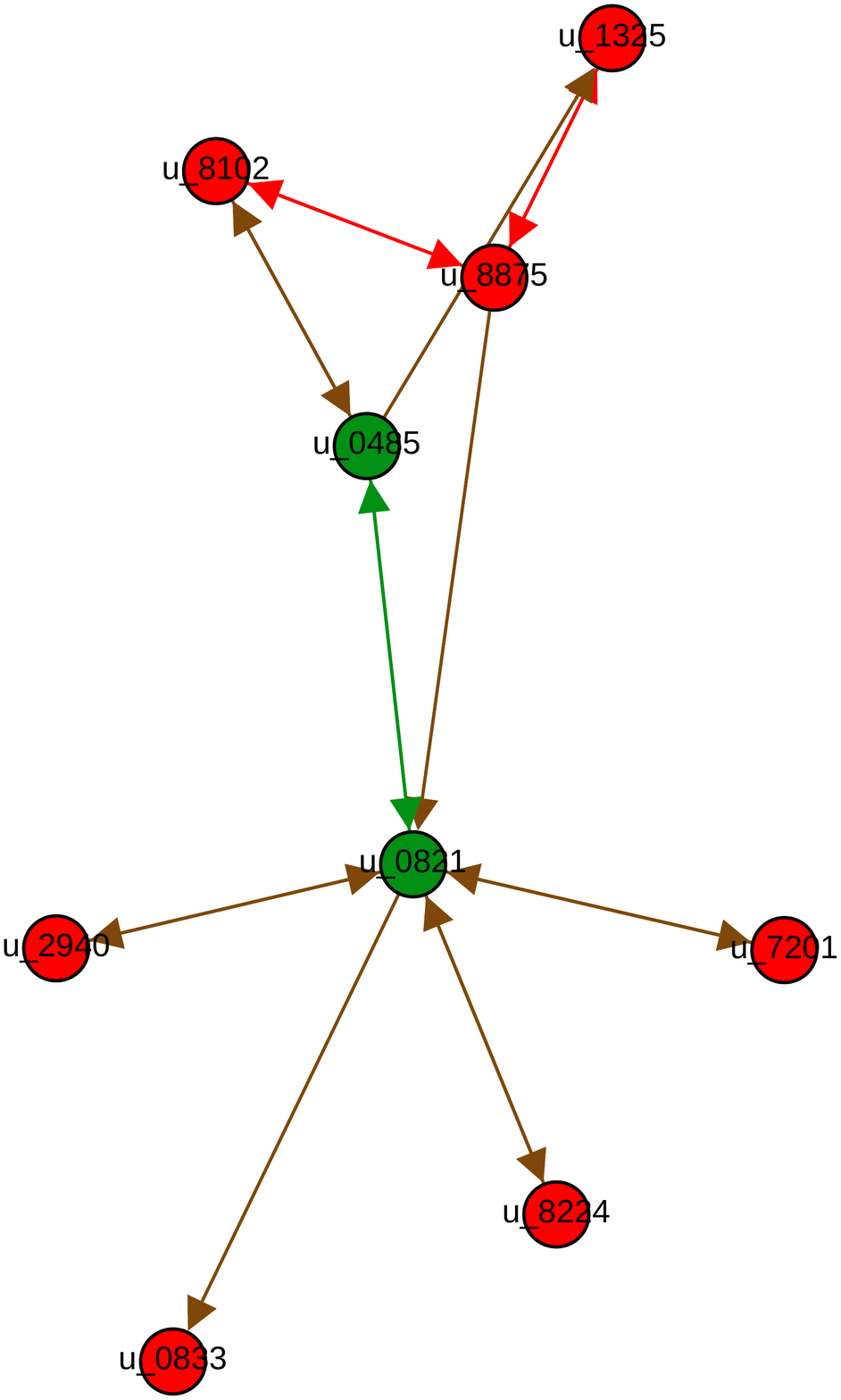}
  }
  \caption{Visualization of fraudulent claims related users over the transaction graph.}
  \label{fig:cases}
\end{figure}

Interestingly, we find that among a cluster of fraudulent users, there are some normal users. To examine this, we choose two typical clusters of fraudulent users and plot their behaviors over the transaction network in Figure~\ref{fig:cases}. 
In the case 1, the fraudulent users (in red) exchange funds through a normal user (in green). This is a typical pattern where fraudulent users do not directly contact, instead, they find a ``normal" user (the exchange hub in the Figure~\ref{fig:cases}a) with a clean record to do so to cover their fraudulent behaviors/monetary traces. 
A similar pattern is also observed in case 2. Differently, we observe some claims between fraudulent users and there are fraudulent gangs connected through two ``normal" users.
In all, user embeddings learned using the transaction graph are insightful and helpful for discovering fraudulent users and claims.

\subsection{Return-Freight Insurance}

Buyers would like to return genuine and undamaged products for various reasons. In some cases, there could be a significant color difference between the on-screen product and the real-life product. In other cases, customers find a less expensive alternative after receiving their goods. The desire to return such items is reasonable but it will raise lots of disputes between buyers and sellers because of the ambiguity over which party should take responsibilities. Most disputes focus on who should pay for return shipping costs. The return-freight insurance is created to resolve disputes and protect buyers' right to regret.

\subsubsection{Graph Comparison}

We analyze the patterns of fraudulent claims in the scenario of return-freight insurance and organized fraud turns out to be the prominent form of fraud. Three graphs - device-sharing graph, transaction graph, and the friendship graph are compared according to the label aggregation measure $\eta$, the device-sharing graph fits the best. The conclusion is also shown in~Table~\ref{tab:aggr}.

\begin{table}[!h]
    \centering
    \begin{tabular}{l c c c }
        \toprule
         & hop 1 $eta$ & hop 2 $eta$ & overall $eta$  \\
        \midrule
        Device-sharing   & 0.80 & 0.51 & 0.80   \\
        Transaction   & 0.16  & 0.06 & 0.16  \\ 
        Friendship   & 0.04 & 0.01 & 0.04   \\
        \bottomrule
    \end{tabular}
    \caption{Label aggregation comparison in terms of graph choice.}
    \label{tab:aggr}
\end{table}

\subsubsection{Data Preparation}

Our return-freight insurance dataset is sampled from its claim history from the past three months. 
The device-sharing graph is constructed with accounts that have filed a claim within a 30-day period. Device UMIDs used by these accounts in the past 40 days are added as graph nodes. Isolated subgraphs containing only one account node are removed for computation efficiency. For raw features of account nodes, we collect 50 features (e.g., number of claims submitted over a month, duration as a customer, etc.), derived from insurance claim history, shipping history, and shopping history. 

\subsubsection{Quantitative Evaluation}

After choosing the proper graph, we compare the DeepWalk algorithm, the GeniePath algorithm, and PSMART. We set the same hyperparameters for all PSMART modules: 500 trees, max tree depth of 5, data sampling rate of 0.6, feature sampling rate of 0.7, and a learning rate of 0.009. We randomly sample 25\% of `no observable risk' accounts as negative samples.

\begin{figure}
\centering
\includegraphics[width=0.9\columnwidth]{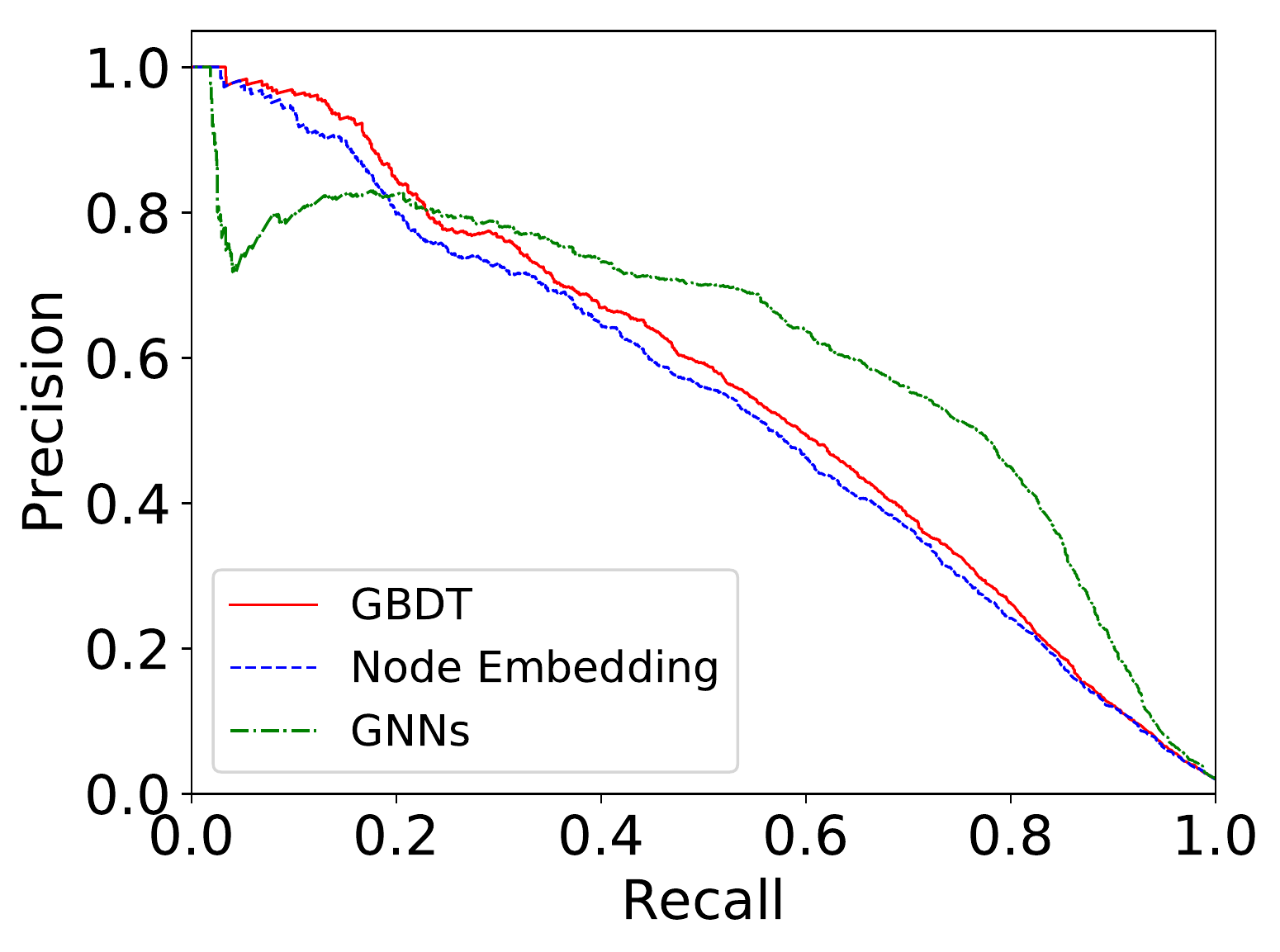}
\caption{Model comparison with the Precision-Recall curve.}
\label{fig:pr}
\end{figure}

Our results, summarized in Table~\ref{tab:results} and plotted in Figure~\ref{fig:pr}, show that the GNNs-based approach outperforms the others. 
Detection expansion (DE), defined as $\frac{FP + TP + FN}{TP + FN}$, indicates the ability to detect more fraudulent accounts.
All of our approaches raise the coverage of fraudulent account detection by more than 40\% while GNNs-based approach has higher precision and recall at most time.

\begin{table}
\centering
  \caption{Results based on Rule-based Labels.}
  \label{tab:results}
  \begin{tabular}{ccccc}
    \toprule
     &  & PSMART & Node Embedding & GNNs \\
    \midrule
    \multirow{ 2}{*}{} & F1 & 0.547 & 0.535 & 0.623 \\
                             & DE & 1.47 & 1.44 & 1.44 \\
  \bottomrule
\end{tabular}
\end{table}

\subsubsection{Online Performance}
Our system collects accounts that have filed a claim over the past months and classifies them daily. The classification result is evaluated by an insurance professional, who randomly samples and examines 300 accounts out of the reported fraudulent accounts. Recent reports show we have a precision of over 80\% while covering 44\% more suspicious accounts.

\subsection{More applications}

\subsubsection{Order Insurance}

The order insurance is generally designed for the same purpose as the security deposit insurance is designed for.  An order insurance policy only covers the lifecycle within one order, and a security deposit policy covers all orders for a specific seller. However, the advanced compensation offered by the insurer is ten times higher. In some categories on Taobao, alcohol, for example, purchasing order insurance is a must for `trustworthy seller' badge since the products cost a large amount of money so the compensation is expected to be higher by the buyers.

By examining the fraudulent claims, we find suspicious relations between some certain buyers and sellers. With the help of the buyer-seller graph and the edge classification algorithm DistRep, recall reaches 89\% in offline experiments. In the online setting, the order insurance using the InfDetect system halves its compensations and saves 
\cc{tens of thousands of dollars per day.}

\subsubsection{Complementary Health Insurance}
Complementary health insurance is offered to buyers as a marketing strategy to foster online shopping activities. 
PSMART is applied with the help of InfDetect and the top 50 suspicious claims are sent to insurance professionals for further investigation. In this specific insurance, human investigation is easier by asking the claimed hospitals for detailed information.  The feedback is not ready yet and more and more other types of insurance are using our system for general fraud detection and organized fraudsters detection.

\section{Related Work}
Traditional methods on \textit{insurance fraud detection} primarily focus on extracting handcrafted features (such as past claim history) and subsequently heuristics/rules are distilled based on expert knowledge to decide whether a claim needs further human investigation or not. Witnessing the emergence of big data and distributed computing, insurance companies have started leveraging machine learning techniques to lessen the burden of human investigation/intervention in the claim process~\cite{sithic2013survey}. Insurance fraud detection approaches can be generally divided into supervised learning, unsupervised learning, and a mixture of both~\cite{joudaki2015using, li2008survey, viaene2002comparison}. 
Popular supervised algorithms, such as logistic regression~\cite{mercer1990fraud, wilson2009analytical}, decision trees~\cite{bonchi1999classification}, support vector machine, Bayesian networks~\cite{ormerod2003using}, and neural networks~\cite{shapiro2002merging, he1997application}, have demonstrated good performances, however, they require data to be labeled by domain experts. 
Meanwhile, unsupervised techniques, such as association rules, cluster analysis, and outlier detection have also been applied and attracted much attention over the years~\cite{brockett2002fraud, yamanishi2004line, viveros1996applying, nian2016auto}.
Hybrids of supervised and unsupervised algorithms have been studied, and unsupervised approaches have been used to segment insurance data into clusters for supervised approaches in~\cite{brockett1998using}.
Our proposed approaches/system fall under supervised learning and hybrids
of both unsupervised and supervised, respectively. Our proposed approaches/system differ, as we are the first to introduce/incorporate graph information into the insurance fraud modeling.

Graph/network provides straightforward information for describing and modeling complex relations among colluders (collaborating fraudsters). It is the most natural representations of relation information and allows for complex analysis without simplification of data. Recently, network representation learning is playing an increasingly important role in network analysis.
Many unsupervised models have been introduced over the years, e.g., the widely used LINE~\cite{tang2015line}, DeepWalk ~\cite{perozzi2014deepwalk}, and node2vec~\cite{grover2016node2vec}, which demonstrated to be superior compared to the traditional graph analysis approaches such as spectral clustering~\cite{tang2011leveraging}, modularity analysis~\cite{tang2009relational}.
Meanwhile, Graph Neural Networks (GNNs) represent a set of supervised graph learning algorithms following the same architecture that aggregates information from nodes' neighbors~\cite{gori2005new, scarselli2008graph}. Commonly used state-of-the-art GNN-based approaches include struct2vec~\cite{dai2016discriminative}, GAT~\cite{velivckovic2017graph}, GeniePath~\cite{liu2018geniepath}, which have demonstrated to be effective in various applications~\cite{liu2018heterogeneous,hu2019cash}.

\section{Conclusion}
In this work, we present a graph-based fraud detection system for large scale e-commerce insurance with the cases of the most popular insurance - the security deposit insurance and the return-freight insurance. We also introduce the modules and their functionality in this system. The key component - graphs and their learning algorithms help discover organized fraudsters and the system 
\cc{has helped save millions of dollars per year}.

\bibliographystyle{IEEEtran}
\bibliography{IEEEexample}

\end{document}